\documentclass[aip,rsi,amsmath,amssymb,reprint]{revtex4-1}
\usepackage{amssymb}
\usepackage{amsmath}
\usepackage{dcolumn}
\usepackage{bm}
\usepackage{color}
\usepackage{graphicx}

\begin{document}

\title{Superconducting spin valves controlled by spiral re-orientation in B20-family magnets}

\author{N. G. Pugach}
\affiliation{Skobeltsyn Institute of Nuclear Physics Lomonosov Moscow State University, Leninskie Gory 1(2), 119991 Moscow, Russia}

\affiliation{\mbox{National Research University Higher School of Economics, 101000, Moscow, Russia}}

\author{M. Safonchik}
\affiliation{A. F. Ioffe Physical-Technical Institute,  RU-194021 St. Petersburg, Russia}

\author{T. Champel}
\affiliation{Univ. Grenoble Alpes, CNRS, LPMMC, 38000 Grenoble, France
}

\author{M. E. Zhitomirsky}
\affiliation{Univ. Grenoble Alpes, CEA, INAC-Pheliqs, 38000 Grenoble, France}

\author{E. L\"{a}hderanta}
\affiliation{\mbox{Lappeenranta University of Technology, P.O. Box 20, FI-53851 Lappeenranta, Finland}}

\author{M. Eschrig}
\affiliation{\mbox{Department of Physics, Royal Holloway, University of London, Egham, Surrey TW20 0EX, UK}}

\author{C. Lacroix}
\affiliation{Univ. Grenoble Alpes, CNRS, Institut N\'{e}el, 38000 Grenoble, France}

\begin{abstract}
We propose a superconducting spin-triplet  valve,
which consists of a superconductor and an itinerant magnetic material, with the magnet showing an intrinsic non-collinear
 order characterized by a wave vector
that may be aligned in a few equivalent preferred directions under control of a weak external magnetic field.
Re-orienting the spiral direction allows one to controllably modify long-range spin-triplet superconducting correlations,
leading to spin-valve switching behavior. Our results indicate that the spin-valve effect may be noticeable.
This bilayer may be used as a magnetic memory element for cryogenic nanoelectronics.
It has the following advantages in comparison to superconducting spin valves proposed previously: (i) it contains only one magnetic layer, which may be more easily fabricated and controlled; (ii) its ground states are separated by a potential barrier, which solves the ``half-select'' problem of the addressed switch of
memory elements.
\end{abstract}
\date \today
\maketitle


{\em Introduction.} Superconducting spintronics is a  
field within nanoelectronics of quantum
systems, which has emerged and is actively developing in the past 20 years. Its main idea is the usage of electronic spin transfer for
information storage and processing, such as usual spintronics, but it is implemented
in superconducting circuits at low temperatures \cite{Eschrig2011,Linder2015,Eschrig15b}.
Among the basic units of superconducting spintronics are the so-called superconducting
spin valves (SSV). These are nano-devices in which the
superconducting current is controlled through the spin degree of freedom, by changing the magnetization of magnetic elements. SSVs may serve as control units of low temperature
nanoelectronics and  magnetic memory elements for spintronics and
low-power electronics \cite{Holmes2013}.

SSVs were theoretically proposed almost 20 years ago
\cite{Beasley1997,BuzdinVed1999,Tagirov1999}  as elements consisting
of a thin  superconducting (S) layer (assuming singlet pairing) and two ferromagnetic (F) layers. Their state is switched from the superconducting to the normal conducting (N) one depending
on the mutual orientation (parallel or antiparallel) of the F layer's
magnetizations, analogously to usual spin valves.
The SSV mechanism is based on
the suppression of the superconducting critical temperature $T_c$ by the magnetic exchange in the
F layers, which influences the S properties via the proximity effect. The two F
layers effectively act together (at parallel magnetization alignment) or at odds
(antiparallel alignment) in the process of reducing superconductivity. Since the superconducting correlations are present on length scales which are larger compared to
atomic scales, SSV, in contrast to usual magnetic spin valves,
can be made in two configurations: SFF \cite{Beasley1997} or FSF \cite{BuzdinVed1999,Tagirov1999}, where the F layers are located at one or at the two sides of the S
layer. It was shown later  \cite{Fominov2010,Aarts2015,Garifullin2013}
 that the SFF configuration is preferable,
because it provides the closest interaction between the two $F$ layers.

The here proposed SSVs take advantage of another physical mechanism.
It was demonstrated 
theoretically \cite{BVE:PRB2001,Shekhter,Eschrig03,VolkovRevModPhys} that a non-collinear magnetization in the SF heterostructures also generates spin-triplet
superconducting correlations with a non-zero 
spin projection on the
quantization axis. The exchange field does not suppress the equal-spin
triplet pairs, which thus penetrate far in the F region.
These long-range triplet correlations (LRTC) were 
revealed experimentally with
 the observation of
long-range superconducting currents in Josephson spin-valves \cite{KeiserNature,Khaire,RobinsonScience}.
 The appearance of LRTC affects the proximity effect by opening a
new channel for the Cooper-pair drainage from the S layer.
In SSVs, the influence of the LRTC on $T_c$ has already been reported in numerous experiments \cite{Westerholt2005,Leksin2012,Zdravkov2013,Krasnov2014,Blamire2014,Blamire2014b,USA2014,Flokstra2015,Aarts2015}.
A direct relationship between the production of spin-polarized triplet correlations and the suppression of $T_c$ has been observed in SSVs under various peculiar conditions \cite{USA2014,Flokstra2015,Garifullin2016,Garifullin2016b,TagirovKupr2015}, with a stronger $T_c$ reduction found in the case of non-collinear magnetizations  than in collinear configurations (antiparallel or parallel).
Such a device is called a triplet spin valve.
The full switch from the S to N states requires in SSVs a $T_c$ reduction which overcomes the typical width of the superconducting transition and has been achieved only very recently by exploiting the LRTC \cite{Garifullin2016}.

However, it is still an open question whether this type of triplet SSV can be used as
switchable elements in real devices. Indeed, in these nanostructures an additional
antiferromagnetic layer is required for the pinning of one F layer by the
magnetic exchange coupling, whereas the magnetization of the second
F layer can be rotated freely. Nonmagnetic layers are also required to separate and decouple the two F layers,
and additional Cu layers are introduced in the intermetallic interfaces to
improve their quality \cite{Garifullin2016b}.  Thus, such a triplet SSV  contains several layers of different  magnetic, nonmagnetic, and antiferromagnetic materials, which is highly demanding for technology and magnetic configuration controlling.

\begin{figure}[t]
\begin{center}
\includegraphics[width=0.4\textwidth]{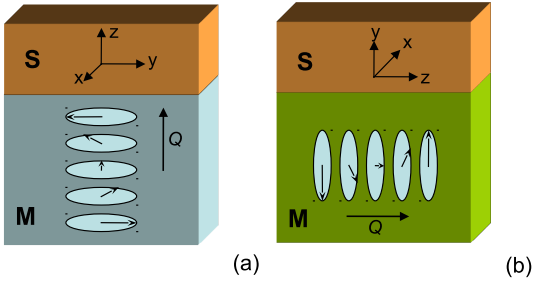}
\end{center}
\caption{(Color online) The sketches of the spiral SSV in two configurations: spiral vector ${\bf Q}$ (a)  orthogonal to the superconducting interface (opening of the LRTC channel),
or (b) parallel to the superconducting interface.}
\label{Sketches}
\end{figure}

In this paper, we consider the realization of a different type of triplet SSV, which contains instead only one magnetic layer with controllable
intrinsic non-collinear magnetization. Suitable magnetic materials can be found in the B20 family of itinerant cubic helimagnets, MnSi, (Fe,Co)Si, and FeGe \cite{Ishikawa1977,Pfleiderer2001,Uchida2006}.
Currently, these compounds and their films are intensively investigated \cite{Muhlbauer2009,Fert2013} as a medium for magnetic topological defects such as skyrmions. Their spiral magnetic
structure characterized by the vector ${\bf Q}$  may be aligned in a few equivalent directions under the control of a weak external magnetic field.
Importantly, these compounds are metals and are thus susceptible to sustain the proximity effect with a superconducting layer.
As shown in Refs. \cite{Champel2005,Champel08}, a superconducting-magnetic spiral (M) bilayer with a spiral vector ${\bf Q}$ parallel to the interface does not generate LRTC components, while the latter are expected when ${\bf Q}$ is inclined with respect to the interface \cite{BVE:PRB2001,Shekhter,VolkovRevModPhys}.
 As illustrated in Fig. \ref{Sketches}, we suggest that the switch between the two spiral order directions
thus controls the opening of a new channel for Cooper pairs drainage
related to the LRTC creation. In the remainder of this paper, we
study quantitatively the change in  $T_{c}$ induced by the magnetic switch and discuss the properties of the proposed bilayer SSV.

{\em Calculation method.}
For simplicity, we consider a S layer with a finite
thickness $d_{s}$ covering a semi-infinite M layer with the
vector $\bf{ Q}$, 
along the $OZ$ direction, which may be parallel or
orthogonal with respect to the interlayer interface (see Fig. \ref{Sketches}).
For the  $T_c$ calculations, we assume the diffusive limit, because real
superconducting nanostructures made by sputtering are usually dirty. In this case, the superconducting coherence lengths in the M and S layers are
given by $\xi _{f,s}=\sqrt{D_{f,s}/2\pi T_{cb}}$, where $D_{f,s}$ is the corresponding diffusion
coefficient and $T_{cb}$ is the critical temperature of the bulk
superconductor.  Close to $T_{c}$, superconducting correlations are weak, so that we can describe
the underlying physics in the framework of the linearized Usadel equations.

The triplet proximity effect in the presence of a spiral or conical magnet
has been intensively studied both theoretically in either the  dirty \cite{Champel2005,VolkovAnishchanka,Halasz2009,Halasz2011}  or the clean \cite{Wu2012,USA2014,Halterman2016} regimes and experimentally \cite{RobinsonScience,Sosnin2006,Zhu2013,Blamire2014b,Blamire2015,Aarts2015,Chiodi2013,DiBernardo2015,Satchell2017} in complex multilayered magnetic structures designed for revealing LRTC.
The theoretical work in Refs.~\cite{VolkovAnishchanka,Halasz2009}
treated the critical current in S-M-S Josephson
junctions with a conical vector orthogonal to the interface. Ref.
\onlinecite{Halasz2009} assumed the limit of a short spiral wavelength $\lambda \ll \xi _{f}$, which is
suitable for Ho but not for MnSi. The case of a long
wavelength was considered in Ref.
\onlinecite{VolkovAnishchanka} with a helicoidal magnet model, but the effect on $T_c$
was not investigated.
The $T_c$ calculation in the parallel configuration (b) as depicted in Fig. \ref{Sketches}  was published
in Ref.  \cite{Champel2005}, which only considered changes in $T_c$ with respect to the amplitude of
${\bf Q}$, not with respect to its direction (which is the topic of this paper).

The linearized Usadel equations for the singlet $f_{s}$ and triplet $%
{\bf f}_{t}=(f_{x},f_{y,}f_{z})$ spin components of the anomalous
Green's function, describing superconducting correlations, have the following form:
\cite{Champel2005}
\begin{eqnarray}
\left( D_{f,s}\nabla ^{2}-2\left\vert \omega \right\vert \right) f_{s}
&=&-2\pi \Delta +2i\ \mathrm{sgn}(\omega ) {\bf h}\cdot {\bf f}_{t}, \label{Uo}
 \\
\left( D_{f,s}\nabla ^{2}-2\left\vert \omega \right\vert \right) {\bf
f}_{t} &=&2i\ \mathrm{sgn}(\omega ){\bf h}f_{s}.  \notag
\end{eqnarray}
The singlet
superconducting order parameter $\Delta $ is
nonzero only in the S layer, while the exchange field $
{\bf h}=h(\cos Qz,\sin Qz,0)$ is nonzero and aligned along the local
magnetization in the M layer. The spiral vector ${\bf Q}$ is always taken
parallel to $OZ$ axis. Since $h_{z}=0$, the third triplet component $f_{z}=0$.
Using the unitary transformation $f_{\pm}=\left( \mp f_{x}+if_{y}\right) \exp
(\pm iQz)$  and taking into
account the symmetry of the Green's functions with respect to the Matsubara
frequency $\omega \equiv \omega _{n}=\pi T(2n+1)$
with $n$ an integer, we may rewrite Eqs. \eqref{Uo} for $\omega \geq 0 $ as
\begin{eqnarray}
\left( D_{f,s}\nabla ^{2}-2\omega \right) f_{s} =-2\pi \Delta +i\ h\left(
f_{-}-f_{+}\right) , \hspace*{1.5cm}  \label{UEq}\\
\left( D_{f,s}\nabla ^{2}\mp 2iD_{f,s}Q \partial_{ z}%
-D_{f,s} Q^{2}-2\omega \right) f_{\pm} =\mp 2i\ hf_{s}.  \notag
\end{eqnarray}
Eqs. (\ref{UEq}) are supplemented by boundary
conditions \cite{KL} at $r=0$, where the coordinate $r=z$ or $r=y$
refers to the distance from the S/M interface depending on the chosen spiral configuration
\begin{equation}
\xi _{s}\partial_{r}f_{s,x,y}^{S}=\gamma \xi_f \partial_{ r}f_{s,x,y}, \,\,
f_{s,x,y}^{S}=f_{s,x,y}-\gamma _{b} \xi_f \partial_{ r}f_{s,x,y},  \label{KLBC}
\end{equation}
with the dimensionless interface parameters $\gamma _{b}=R_{b}A\sigma _{f}/\xi_f$ and
$\gamma =(\sigma _{f}/\sigma _{s})(\xi _{s}/\xi_f)$ ($R_{b}$ and $A$ are 
the resistance and the area of the S-M interface, respectively, and $\sigma _{f,s}$ is the
conductivity of the M or S metal).
Boundary conditions (\ref{KLBC})  relate the superconducting correlations coming from each side of the interface.
The correlations in the S layer (located at $r<0$) are described by the functions $f_{s,x,y}^{S}$, while
the functions $f_{s,x,y}$ contain the information about the structure of the M layer ($r>0$).

One obtains via some straightforward algebra 
(see supplementary material)
a closed boundary value problem for the singlet component $f_{s}^{S}$ with the following
boundary condition at the S/M interface:
$\left.\xi _{s}\partial_{r}f_{s}^{S}\right|_{r=0}=- \left.Wf_{s}^{S} \right|_{r=0}$.
The proximity effect in the M layer is entirely encapsulated in the real-valued quantity $W$.
This key-quantity is the subject of the subsequent calculations considering two different spiral alignments in the M layer.
The triplet solutions of Eqs. (\ref{UEq}) 
in the S layer $f_{x,y}^{S}$
may be found in a simple exponential form with the wave vector $k_s=\sqrt{Q^2+2|\omega|/D_s}$. 
Eqs. (\ref{UEq})  for the singlet component include the
coordinate-dependent $\Delta $, which should be calculated
self-consistently. The transition temperature of the structure,
$T_c$, is computed numerically from
\begin{equation}
\ln \frac{T_{cb}}{T_{c}}=\pi T_{c}\sum_{\omega =-\infty }^{\infty } \left(\frac{1}{
\left\vert \omega \right\vert} -\frac{f_{s}^{S}}{\pi \Delta } \right)
\label{SelfCons}
\end{equation}
by using the method of fundamental solution \cite{Fominov2002,Champel2005,Lofwander07}.

{\em Configuration (a): The spiral vector is orthogonal to the S layer.}
In this case, the problem becomes one-dimensional as in Ref. \onlinecite{VolkovAnishchanka} with $r=z$ and
$\nabla ^{2}=\partial^{2}_{ z}$. The M layer is infinite, so that the
functions $f_{s,x,y}$ in the magnetic layer are sought under the form of decaying exponents $\exp
(-kz)$ with $k$ a wave vector determined from the characteristic equation of the linear system (\ref{UEq})
\begin{eqnarray}
\left[ (k^{2}-k_{\omega }^{2}-Q^{2})^{2}+4Q^{2}k^{2}\right] \left( k^{2}-k_{\omega }^{2}\right)
\notag \\
+4k_{h}^{4}(k^{2}-k_{\omega }^{2}-Q^{2})=0,
\label{CharOrt}
\end{eqnarray}
where $k_{\omega }^{2}=2\omega /D_{f}$ and $k_{h}^{2}=h/D_{f}$. Provided that
$k_{h}^{2}\gg Q^{2}>k_{\omega }^{2}$, Eq. \eqref{CharOrt}
yields 3 complex-valued eigenvalues: two solutions $~k_{\pm }=\left( 1\pm i\right) k_{h}$ of the order of $k_{h}$ describing
short-range correlations and one real solution $k_{0}=\sqrt{k_{\omega }^{2}+Q^{2}}$ of the order
of  $Q$ characterizing long-range correlations. The reflected  short- or long-range waves
are neglected, so the M layer should be thicker than $max{[\lambda/\pi,2\xi_f]}$
to be considered as semi-infinite.
Boundary conditions (\ref{KLBC}) together with triplet solutions of Eqs. (\ref{UEq})
in the exponential form with the wave vectors $k_s$ in the S layer
yield  the quantity $W$ when the vector ${\bf Q}$ is orthogonal to the S layer:
\begin{equation}
W_{\perp}=\frac{\gamma \mathrm{Re}\left\{ \xi _{f}k_{+}\left[ (1+\xi_{c}k_{-})\left( 1+\xi_{c}k_{0}\right) -(\xi _{c}Q)^{2}\right] \right\} }{\mathrm{Re}\{\left( 1+\gamma _{b}\xi
_{f}k_{+}\right) \left[ (1+\xi _{c}k_{-})\left( 1+\xi_{c}k_{0}\right) -(\xi_{c}Q)^{2}\right] \}},  \label{Wort}
\end{equation}
with the length $\xi _{c}\equiv \xi_f \left[\gamma \coth
(k_{s}d_{s})/k_{s}\xi _{s}+\gamma _{b} \right]$. The terms containing $Q$ and $k_{0}$
typically characterize the contribution of the long-range triplet correlations.

{\em Configuration (b): The spiral vector is parallel to the S layer.}
We assume that the interface coincides with the $XOZ$ plane.
Since the structure breaks translation invariance
in the $r=y$ direction and the magnetic structure is nonuniform in the $z$ direction, the kinetic
energy operator now reads $\nabla ^{2}=\partial^{2}_{y}+\partial^2_{z}$.
It has been shown in Ref. \onlinecite{Champel2005} that $z$-independent correlations $f_{s,\pm }$
provide the lowest $T_{c}$ (i.e., they are the most favorable energetically),
because they are the only solutions which realize a spatially homogeneous superconducting order
parameter in the S layer at large distance from the S/M interface. In this case, we have the triplet
vector ${\bf f}_{t} || {\bf h}$, meaning that only short-range
triplet components are present.

More precisely, solutions of Eq. (\ref{UEq}) in the M layer  have again the form
of decaying exponents with the wave vectors $k$ determined by a characteristic
equation, with, however, an expresion now simpler than given by Eq. (\ref{CharOrt}),
\begin{equation}
(k^{2}-k_{\omega }^{2}-Q^{2})^{2}(k^{2}-k_{\omega
}^{2})+4k_{h}^{4}(k^{2}-k_{\omega }^{2}-Q^2)=0.  \label{CharPar}
\end{equation}
This equation has an exact solution with two short-range eigenvalues $\tilde{k}_{\pm }=%
\sqrt{2k_{\omega }^{2}+Q^{2}/2\pm (i/2)\sqrt{16k_{h}^{4}-Q^{4}%
}}$ and a long-range one $k_{0}=\sqrt{k_{\omega }^{2}+Q^{2}}.$
Under the
inequalities $k_{h}^{2}\gg Q^{2}$ and $k_{\omega }^{2}$, one gets
$\tilde{k}_{\pm} \approx k_{\pm }$, so that
respective eigenvectors coincide with those obtained for the case of
the orthogonal spiral. 
Thus, under these conditions, the general
form of the superconducting correlations in the M layer in the parallel configuration is identical to that in the orthogonal
configuration. The major difference takes place only in the boundary conditions:
since $\partial_{y}\exp (\pm iQz)=0$, the LRTC with the eigenvector
$k_{0}$\ now has no source neither in the S nor in the M layers, nor at the
interface. 
Following the same steps as in the configuration (a),
we get after straightforward algebra the quantity $W$ when
${\bf Q}$ is parallel to the S layer:
\begin{equation}
W_{\parallel}=\frac{\gamma \mathrm{Re}\left[ \xi _{f}\tilde{k}_{+}\left( 1+\xi_{c}\tilde{k}_{-}\right) \left( \sqrt{16k_{h}^{4}-Q^{4}}+iQ^{2}\right) \right] }{
\mathrm{Re}\left[ \left(1+\gamma _{b}\xi _{f}\tilde{k}_{+} \right)\left( 1+\xi_{c}\tilde{k}_{-}\right) \left( \sqrt{16k_{h}^{4}-Q^{4}}+iQ^{2}\right) \right] }.
\label{Wpar}
\end{equation}
In contrast to $W_{\perp}$, we note that $W_{\parallel}$ does not contain  $k_0$ characterizing the long-range triplet correlations.

{\em Numerical results and discussion.}
For the calculations, we have assumed a S layer made of Nb.
Bulk Nb has a critical temperature $T_{cb}= 9.2$ K. Other data about Nb needed for the calculations
are taken from the experimental work \cite{Flokstra2015}, where, for instance, $\xi _{s}=11$ nm.
Ho magnets displaying conical spiral order have already been used in several hybrid structures
\cite{RobinsonScience,Sosnin2006,Blamire2015} in combination with Nb. The 
control of the superconducting state by a change of the magnetic state has even been obtained
in recent experiments \cite{DiBernardo2015,Satchell2017}. However, the helimagnets Ho or Er
used in these experiments have a strong in-plane magnetic anisotropy
(with a spiral vector remaining orthogonal to the layer plane), so they do not appear as the best choice
for the proposed SSV. Furthermore, an increase in the temperature of the sample
above the Curie temperature (much higher than $T_c$) is needed in order to return
to the initial helimagnetic state \cite{Satchell2017}, which makes this kind of
switch difficult to use in low-temperature electronics.

In contrast, the transition-metal compounds of the MnSi family have a weak magnetic anisotropy
(much smaller than the exchange energy $h$). They crystallize in a noncentrosymmetric cubic B20 structure,
allowing a linear gradient invariant \cite{Bak1980}. This gives rise to a long-period spiral magnetic structure ($\lambda=18$ nm for MnSi).
 The existence of a domain structure with different spiral directions in neighboring domains was
observed in such compounds \cite{Grigoriev2006b}. Importantly, the spiral direction may be
switched in not too large magnetic fields \cite{Grigoriev2006}.
In MnSi, the spiral wave vector ${\bf Q}$ is aligned along [111] and equivalent directions of the cubic lattice.
The angle between these directions is $\alpha =\arccos (1/3)=70.5^{\circ }.$
If one spiral axis is parallel to the S layer plane and another one is inclined by $\alpha$ with respect to this plane,
then
the period of the in-plane magnetic inhomogeneity is $\lambda / \cos \alpha =3\lambda \gg \xi_s$,
which allows neglecting the in-plane inhomogeneity. In Eqs. (\ref{Wort}) and (\ref{Wpar}), the value $Q$
changes to $Q \sin\alpha \approx 0.94 Q$, and the behavior is practically the same as for $\alpha = 90^{\circ }$.
Considering transport properties
\cite{Lee2007,Jonietz2010}, we used the following parameters for diffusive MnSi:
$\xi _{f}=4.2$ nm and 
$\gamma =0.7$. 
The other estimations are $h \sim 100$ meV that is much less than the Fermi energy $\sim 1$ eV, and
$Q=2\pi /\lambda \approx 0.35 \, {\rm nm}^{-1}$. As a result, one gets $%
k_{h}\sim 0.7 \, {\rm nm}^{-1}$  and $k_{\omega }\sim
0.14 \, {\rm nm}^{-1}$. The inequalities $k_{h}>Q>k_{\omega }$ are fulfilled,
thus justifying the approximations made in the derivation of  quantity $W$.

\begin{figure}[!t]
\begin{center}
\includegraphics[width=0.45\textwidth]{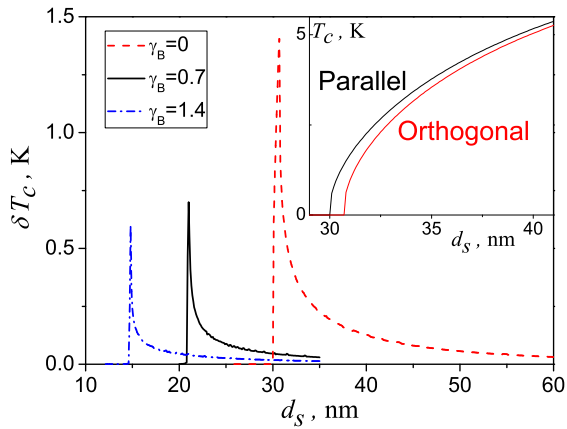}
\end{center}
\caption{(Color online) Change in the superconducting critical temperature $\delta T_c$ at the crossover between the two spiral configurations shown in Fig. 1
as a function of the S layer thickness $d_s$ for different $\gamma_b$ values.}
\label{Fig2}
\end{figure}

The numerical results for $T_c$ are displayed in Fig. \ref{Fig2}.
The magnetic switch from the parallel to the inclined configuration
creates the condition for the LRTC appearance in the superconducting spin valve. As clearly seen in the inset of Fig. \ref{Fig2}  showing an ideal situation for the SSV where the interlayer resistance $\gamma_b$ is neglected, the related drainage of Cooper
pairs from the S layer effectively increases the proximity effect and suppresses $T_{c}$.
The difference $\delta T_{c}$  between the two $T_c$ obtained in each magnetic
configuration is shown in the main panel of Fig. \ref{Fig2} for different values of $\gamma_B$.
It increases when the thickness of the S layer approaches
the critical thickness corresponding to a total disappearance of superconductivity.
Naturally, at this thickness, the superconducting film is most sensitive to the
proximity effect.
The  two left 
curves  account for a more realistic case with
$\gamma _{b}=0.7$, calculated according to Eq. (A14) of Ref. \onlinecite{PugachPRB2011}, and a twice larger $\gamma _{b}=1.4$, assuming an additional interface tunnel barrier.
The experimental value of the coupling between Nb and a weak ferromagnet \cite{RyazanovPRL2006} was $\gamma _{b}=0.5$.
As expected, the consequence of $\gamma_b$ is to weaken the superconducting proximity effect.
The maximum value of  $\delta T_{c}\sim 1$ K is obtained at the thickness where the S layer turns to the normal state in the orthogonal configuration. For realistic nonzero $\gamma_b$, the values of  $\delta T_{c}$, which are between one- to a few hundred mK,
may be still noticeable.
This theoretical result indicates a 
spin-valve effect in the S-M bilayer at $\gamma _{b}=0$, of the same order
of magnitude as predicted \cite{Fominov2010} in triplet SFF spin-valves in the same approximation \cite{footnote}.

SSV structures are promising for application as elements of
magnetic memory for low-temperature electronics, which has become recently a
rapidly developing research direction \cite{Baek2014,Birge2014} related to the urgent need for energy-efficient logic for supercomputers.
Usage of only one S layer and a bulk M layer should significantly simplify the
SSV production technology. Another crucial feature of the proposed SSV is as follows:
The switch of a particular memory
element in a random access memory (RAM) device is carried out
by a net
of two crossed arrays of control electrodes. When the recording signal is sent
along two crossed lines, the memory element located at the intersection of
these lines changes its state, while all other element states in the same row or in the same column (thus receiving only one half of the signal) as the selected element remain unperturbed.
Thus, the control signal should be able to switch the  element state, whereas the half-amplitude signal should not cause a switch. The existence of a potential barrier between the two
equilibrium spiral alignments in the magnetic part of the SSV provides an intrinsic
solution to this half-select problem \cite{Vernik2013}. In contrast, this property
is not present in previously studied triplet SSVs based on the continuous
rotation of magnetization of one the ferromagnetic layers.

In conclusion, we have proposed a 
superconducting spin
valve, consisting of a thin superconducting layer covered by a bulk spiral
ferromagnet with multiple equilibrium configurations, such as MnSi. Its principle of operation is based on the controlled manipulation of
long-range spin-triplet superconducting correlation in the structure. Our numerical results indicate
that the spin-valve effect in these structures may 
be noticeable.
Such spin-valves are superior in various regards compared to previously studied spin-valve geometries.

\vspace*{0.3cm}
See supplementary material for a detailed calculation of the closed boundary conditions for the singlet component of the anomalous Green's function in the S layer.

\vspace*{0.3cm}
This work benefited from
the Visiting Scientist Program of the Centre de Physique Th\'{e}orique de Grenoble-Alpes
and from the Royal Society International Program (IE150246). Also,
N.P. thanks the Project N T3-89 ``Macroscopic quantum phenomena at low and ultralow temperatures'' of the National Research University Higher School of Economics, Russia.

\pagebreak



\widetext

\section*{Supplementary Material: Calculation of the closed boundary conditions for the S layer}\label{S}

To obtain the quantity $W$ in the closed boundary conditions  $\xi_{s} \partial_{r} \left. f_s^{S} \right|_{r=0}= - W \left. f_s^{S} \right|_{r=0}$, 
we should recalculate it from the solution of the boundary-value problem for the triplet components in the S and M materials.
This quantity $W$ entirely describes the proximity effect of the M layer.

The $r$-dependence of the triplet components in the S layer should also obey to the
boundary condition at the free S interface

\begin{equation}
\frac{\partial }{\partial r} f_{s,x,y}^{S}(-d_{s})=0 \label{FreeS}.\eqno(\mathrm{S}1) \nonumber
\end{equation}
The solution of the Usadel equations (2) 
satisfying the boundary condition (S1)
may be written in the form:

\begin{equation}
f_{x}^{S}=C_{x}\frac{\cosh \left[k_{s}(r+d_{s})\right]}{\sinh \left[k_{s}d_{s}\right]},
\eqno(\mathrm{S}2)\nonumber
\end{equation}
\begin{equation}
if_{y}^{S}=C_{y}\frac{\cosh \left[k_{s}(r+d_{s})\right]}{\sinh \left[ k_{s}d_{s}\right]}, \nonumber
\label{S:f_xy}
\end{equation}
where $k_s=\sqrt{Q^2+2|\omega|/D_s}$.
The coefficients $C_{x}$ and $C_{y}$ are found from the boundary conditions (3).

In the case of the spiral vector orthogonal to the S layer $r=z$,
the characteristic equation (5) 
yields 3 eigenvalues $k_{\pm,0 }$.
The corresponding eigenvectors  are $(-1,-1,1),~(1,-1,1)$ for
$k_{\pm }$, and $(0,1,1)$ for $k_{0}$.
These results entirely coincide with those obtained in Ref. [36].

Therefore, considering the above inequalities, the solution of coupled Eqs. (2)
reads in the M layer:

\begin{equation}
f_{s}(z) = -A_{1}\exp (-k_{+}z)+A_{2}\exp (-k_{-}z),\eqno(\mathrm{S}3) \nonumber  \label{fformOrt}
\end{equation}
\begin{equation}
f_{\pm}(z) =\pm A_{1}\exp (-k_{+}z)\pm A_{2}\exp (-k_{-}z)+A_{0}\exp (-k_{0}z).\eqno(\mathrm{S}4) \nonumber
\end{equation}

Using
boundary conditions (3) 
and  relations (S2)-(S4) 
for the triplet components, we then get a first set of 4 equations for the
coefficients $A_{0,1,2}$ and $C_{x,y}$
\begin{equation}
\xi _{s}\frac{\partial }{\partial z}f_{x}^{S}
=\gamma \xi _{f}\left[
A_{1}k_{+}+A_{2}k_{-}-iQA_{0}\right],  \label{BCort} \eqno(\mathrm{S}5) \nonumber
\end{equation}
\begin{equation}
\xi _{s}\frac{\partial }{\partial z}f_{y}^{S}
=-\gamma \xi _{f}\left[
A_{0}k_{0}+iQ\left( A_{1}+A_{2}\right) \right], \nonumber
\end{equation}
\begin{equation}
C_{x}\coth k_{s}d_{s}=-A_{1}-A_{2}-\gamma _{b}\xi _{f}\left(
A_{1}k_{+}+A_{2}k_{-}-iQA_{0}\right), \nonumber
\end{equation}
\begin{equation}
C_{y}\coth k_{s}d_{s} =A_{0}+\gamma _{b}\xi _{f}\left[
A_{0}k_{0}+iQ\left( A_{1}+A_{2}\right) \right] . \nonumber
\end{equation}

By writing  boundary conditions (3) 
for the singlet component, we obtain two additional equations
\begin{eqnarray}
\xi _{s}\frac{\partial }{\partial z}f_{s}^{S}
&=&\gamma \xi _{f}\left(
A_{1}k_{+}-A_{2}k_{-}\right), 
\nonumber
\\
f_{s}^{S}
&=&-A_{1}+A_{2}-\gamma _{b}\xi
_{f}\left( A_{1}k_{+}-A_{2}k_{-}\right),  \nonumber
\end{eqnarray}
which, 
by eliminating the coefficients $A_{0,1,2}$ and $C_{x,y}$,
yield  the quantity $W$  in the orthogonal configuration [Eq. (6)] 
.

The calculation of the quantity $W$ [Eq. (8)] 
for the case where the spiral vector ${\bf Q}$ is parallel to the S layer
is carried out in the same way.


\end{document}